\documentclass[aps,twocolumn,preprintnumbers,%
superscriptaddress,floatfix]{revtex4}
\usepackage[dvips]{graphics}
\usepackage{amsmath,amsfonts,bm}
\usepackage{epsfig}
\begin{document}
\newcommand{\be}{\begin{eqnarray}}
\newcommand{\ee}{\end{eqnarray}}
\def\lsim{\mathrel{\rlap{\lower3pt\hbox{\hskip1pt$\sim$}}
     \raise1pt\hbox{$<$}}} %less than or approx. symbol
\def\gsim{\mathrel{\rlap{\lower3pt\hbox{\hskip1pt$\sim$}}
     \raise1pt\hbox{$>$}}} %greater than or approx. symbol
\def\N{${\cal N}\,\,$}
\newcommand\<{\langle}
\renewcommand\>{\rangle}
\renewcommand\d{\partial}
\newcommand\LambdaQCD{\Lambda_{\textrm{QCD}}}
\newcommand\tr{\mathrm{Tr}\,}
\newcommand\+{\dagger}
\newcommand\g{g_5}

\title{Matter Formed at the BNL Relativistic Heavy Ion Collider}
\author {G.E. Brown}
\affiliation { Department of Physics and Astronomy\\
State University of New York, Stony Brook, NY 11794-3800}
\author {B.A. Gelman}
\affiliation { Department of Physics and Astronomy\\
State University of New York, Stony Brook, NY 11794-3800}
\author{Mannque Rho}
\affiliation{ Service de Physique Th\'eorique,
 CEA Saclay, 91191 Gif-sur-Yvette c\'edex, France and\\
School of Physics, Seoul National University, Seoul 151-747,
Korea}

%\date{\today}
\begin{abstract}
We suggest that the ``new form of matter" found just above $T_c$ by
RHIC is made up of tightly bound quark-antiquark pairs, essentially
32 chirally restored (more precisely, nearly massless) mesons of the
quantum numbers of $\pi$, $\sigma$, $\rho$ and $a_1$. Taking the
results of lattice gauge simulations (LGS) for the color Coulomb
potential from the work of the Bielefeld group and feeding this into
a relativistic two-body code, after modifying the heavy-quark
lattice results so as to include the velocity-velocity interaction,
all ground-state eigenvalues of the 32 mesons go to zero at $T_c$
just as they do from below $T_c$ as predicted by the vector
manifestation (VM in short) of hidden local symmetry. This could
explain the rapid rise in entropy up to $T_c$ found in LGS
calculations. We argue that how the dynamics work can be understood
from the behavior of the hard and soft glue.
\end{abstract}
\maketitle
% \vspace{0.1in} ]
%\begin{narrowtext}
%\newpage

\date{\today}

\newcommand\sect[1]{\emph{#1}---}

\maketitle

\sect{Introduction} Within a month or two of operation, RHIC found
the ``new matter" they formed to be an extremely strongly
interacting liquid, nothing like the quark gluon plasma.

This brings up two issues: a long-standing ``old" issue as to what
the structure of the state is in the vicinity of, and below, the
presumed chiral phase transition point and what the proper tool to
understand it is and a ``new" issue as to what lies above the
critical point to which the accepted theory of strong
interactions, QCD, is supposed to be able to access
perturbatively. In this Letter, we wish to address these issues in
terms of an old idea on in-medium hadron properties proposed in
1991~\cite{BR} which has been recently modernized with the
powerful notion of ``vector manifestation (VM)" of hidden local
symmetry theory~\cite{HY} and buttressed with results from lattice
QCD. Our principal thesis of this paper is that just as it
required an intricate and subtle mechanism to reach the VM
structure of chiral symmetry just below $T_c$ -- which is yet far
from fully understood -- from the standard linear sigma model
picture applicable (and established) at $T\sim 0$, the structure
of matter infinitesimally above $T_c$ could also be vastly
intricate and subtle from the starting point of QCD at $T\sim
\infty$ at which asymptotic freedom is applicable and
``established." We will make here an admittedly leaping conjecture
by extrapolating and inferring from available information coming
from lattice results and hinted by RHIC data that the structures
of matter just below and just above $T_c$ could be related.
Further arguments to support this conjecture will be presented in
a later publication.

In this paper we shall focus on a new state of matter formed at
high temperature probed by RHIC experiments, particularly in the
vicinity of the critical temperature $T_c\sim 175$ MeV. We should
mention that the question of how matter changes from one form of
symmetry realization to another form of symmetry realization is a
general and fundamental issue in physics and in the case of
hadrons, this still remains a more or less open question. Our
conclusion drawn based on a variety of arguments to be developed
below will be that the transition is ``continuous" across $T_c$ in
the sense that the same degrees of freedom that come up from below
$T_c$ are found just above $T_c$ although perhaps in a disguised
form.

As we will develop in this paper, the key to the possible new
matter produced at RHIC and its connection to the matter below
$T_c$ is in the glue from the gluons exchanged between quarks and
its role in Brown-Rho scaling.

\sect{Two types of glue} The first observation we make is that
there are two types of glue, a soft one and a hard one (epoxy).
This aspect has been emphasized recently in \cite{BLR-twoglues}.
The fact that there are two glues was first found in the thesis
work at Columbia by Yuefan Deng \cite{YD}. He found in lattice
gauge simulations (LGS) that about half of the total glue, the
soft glue, was melted as $T$ reached $T_c$. The exchange of soft
gluons holds hadrons together at temperatures below the phase
transition temperature $T_c = 175 \, MeV$. Below this temperature
their coupling to hadrons produces a soft gluon condensate which
is melted as $T$ goes up to $T_c$. As the soft glue melts the
constituent quarks turn into massless current quarks. The hard
glue (epoxy) on the other hand begins to melt only above $T_c$. It
is the epoxy condensate that produces the length parameter
$\Lambda_{QCD}$ by breaking scale invariance and through what is
called ``dimensional transmutation.''

\sect{Brown-Rho scaling} In 1991 \cite{BR} Brown and Rho predicted
using a dilaton field that as the soft glue melted with increasing
temperature (or increasing density) hadron masses would decrease
essentially proportional to the decrease in the soft gluon
condensate, going to zero at $T_c$. Their scaling found in
\cite{BR}, ${m^{*} \over m} \sim\frac{f_\pi^*}{f_\pi}$, can be
related to the quark condensate,
% ~\cite{footnote1},
\begin{equation}
{m^{*} \over m} \sim {\langle \bar{q} q \rangle^{*} \over \langle
\bar{q} q\rangle} \ .\label{BR}
\end{equation}
Two recent independent studies in a generalized hidden local
symmetry approach, one by Harada and Sasaki~\cite{HS} and the
other by Hidaka et al.~\cite{hidakaetal}, confirmed that both the
vector and axial vector mesons go massless at the chiral
restoration satisfying (\ref{BR}).

 This so-called ``Nambu scaling" was confirmed in heat bath by
Koch and Brown \cite{KB} from lattice results.
%Two of us (G.E. Brown
%and B.A. Gelman) are in the process of updating the Koch and Brown
%work \cite{KB},
%This work needs to be updated since it was based on 14-year old
%lattice gauge calculations.
Thermodynamics of quasiparticles; {\it
i.e.}, that the entropy of a hadron with effective mass $m^{*}$
can be obtained from that of mass $m$, simply replacing $m$ by
$m^{*}$, holds up. So entropy is just counting. The main
correction to the early work is that the $T_c =140 \, MeV$ used
there should be increased to $T_c =175 \, MeV$. The energy in the
gluon condensate is therefore increased by $(140 /175)^{4} =2.44$.
This is important for the bag constant, which is just $1/4$ of the
gluon condensate.

There are experimental indications that vector meson masses do
undergo Brown-Rho scaling in medium. It has been shown by Shuryak
and Brown \cite{SB}that the results of the STAR detector at RHIC
demonstrated that the $\rho$ meson mass was somewhat decreased at
low density at RHIC and a recent CBELSA/TABS collaboration
experiment\cite{rho} unambiguously showed that the mass of the
$\omega$ meson while inside a tin nucleus is also less than in
free space. Both decreases follow the quantitative estimates of
Brown and Rho with Nambu scaling, corresponding to an $\sim 20 \%$
drop in mass by nuclear matter density.

\sect{Meson masses are zero at $T_c$} It was argued by Brown et
al.~\cite{BJBP} that mesons whose masses decrease with temperature
to zero at $T_c$ could describe the large entropy increase found
in lattice gauge calculations as $T \to T_c$. (Massless mesons
give a greater entropy than massive ones.) In other words, the
phase transition was described as mesons going massless, that is,
chirally restored. The spontaneous breaking of chiral symmetry
which gives the mesons their scalar masses is restored at $T_c$.
It is plausible that the multiplet structure at $T_c-\epsilon$
reflects Weinberg's ``mended symmetry"~\cite{weinberg}.

\sect{Thermal masses.} One of the surprises of the Bielefeld
lattice gauge calculations was that they found the quarks to have
very large thermal masses, of the order of $m_q \sim 1 \, GeV$ at
$T=1.5 \, T_c$ and similarly large masses at nearby temperatures.
These thermal masses arise from the self-energy diagrams at finite
$T$ and, in our scenario, are the only masses just above $T_c$. It
should be stressed that they are $not$ scalar masses which break
chiral symmetry but energies, that is, fourth components of four
vectors. It has not been possible to calculate these masses
analytically since the interactions just above $T_c$ are supposed
to be strong and hence nonperturbative. Weldon~\cite{weldon}
obtained in perturbation theory a somewhat complicated formula
involving momentum for the dispersion relation which could however
be approximated to within $\sim 10$ \% by $p_0^2\approx m_q^2 +
\vec{p}^2$ with $m_q$ a temperature dependent quantity which could
be used as an effective mass in a simple hydrogenic model derived
from Bethe-Salpeter equation as done in \cite{PLB} following the
argument of Hund and Pilkhun~\cite{pilkuhn}. Remarkably, despite
the presence of the effective mass, the helicity remains conserved
as the quark wave function satisfies a free Dirac equation. We
shall assume that this result holds non-perturbatively with the
masses given by the Bielefeld lattice calculations.

Now $T_c =175 \, MeV$ so that the Boltzmann factor
%\cite{footnote2}
for the quark $e^{-m_{q}/T}$ is tiny. Similar results were found
for gluons. Thus, there are not enough quarks and gluons present
to produce the pressure, etc. found in RHIC experiments. If the
mesons $\pi$, $\sigma$, $\rho$, $a_{1}$ in $SU(4)$ multiplets go
massless at $T_c$, they can furnish the pressure. But they can
only go massless if the binding of $\sim 1 \, GeV$ quarks and
antiquarks is $\sim 2 \, GeV$, so the bound states have to be
small, $\sim \hbar/(2 \, GeV)$ in radius, much smaller than the
typical $\hbar /(m c)$ of the usual mesons, where $m$ is their
mass. The calculation in \cite{PLB} (described below) found the
rms radius of the bound states just above $T_c$ to be $\lsim 0.4$
fm. The pion is an exception to this. The interaction must bring
$2 m_q \sim 2 \, GeV$ to zero, and therefore is very strong.

The Bielefeld group \cite{b1} have carried out lattice gauge
calculations to obtain the heavy quark free energy and entropy for
the region of temperatures above $T_c$. Their results (for pure
gauge) were used by Park, Lee and Brown \cite{PLB} in agreement
with their full QCD calculations \cite{b3} in almost all respects,
once a rescaling by the relevant temperatures is made. {}From the
energy and entropy the Coulomb potential $V$ can be obtained
\begin{equation}
V(r, T) = F(r, T) - T \, {\partial F(r, T) \over \partial T} \, .
\end{equation}
We shall pay particular attention to $T \gsim T_c$, where we have
$V(r)$, suppressing the $T$. The caveat here is that defining the
thermal modification of a potential energy between the
quark-antiquark pair is known to be complicated and
subtle~\cite{karsch}.

Now the calculations are for quarks of infinite mass. In order to
use them for light quarks, which make up the $SU(4)$ multiplets
$\pi$, $\sigma$, $\rho$, $a_1$ we must put in the magnetic
(velocity dependent) additions. We will be guided by Brown
\cite{GB} who considered the velocity dependence of the
interaction between the two $K$-electrons in heavy atoms. In the
case of Uranium, $Z \alpha \sim 2/3$ so we are not so far from the
$\alpha_s \sim 1$ encountered here. Since the color singlet
interaction is simply a Coulomb one, the difference from the
Coulomb interaction in atomic physics being that since QCD is a
gauge theory, it runs with scale, we can use the apparatus of
high-Z atomic physics to put in the velocity dependence. A nice
result of our procedure is that we obtain the correct number of
degrees of freedom, that of 32 massless bosons as determined by LGS.
The total interaction for stationary states
is, translated to QCD,
\begin{equation}
V = -\, {\alpha_s \over r} \, \left( 1- \vec{\alpha}_1 \cdot
\vec{\alpha}_2 \right) \, ,
\end{equation}
where $\vec{\alpha}_1$ and $\vec{\alpha}_2$ are the Dirac velocity
operators. The $\vec{\alpha}$'s are also helicities, and have
eigenvalues $\pm 1$ above $T_c$. Thus, since the quark and
antiquark move in opposite directions, $\vec{\alpha}_1 \cdot
\vec{\alpha}_2 =-1$ and
\begin{equation}
V = -\, {2 \, \alpha_s \over r} \, .
\end{equation}
These results follow because the quark and antiquark are in
helicity eigenstates~\cite{weldon}.

Lattice gauge calculations above $T_c$ show collective excitations
S, P, V and A, all degenerate \cite{AHN}. These are just the
Nambu-Jona-Lasinio collective states (collective because they are
a sum over quarks and quark-holes) which continue on up through
$T_c$ \cite{BJBP}.

In \cite{PLB} the masses of the bound states (all 32 are
degenerate) above $T_c$ were calculated with input of the lattice
results for the Coulomb potential \cite{b1,b3} with magnetic
effects added. Since, the (thermal) quark masses $m_q$ have not
been calculated at $T_c$ (only at $1.5 T_c$ and $3 T_c$)
calculations were made for various assumptions within the range
$1-2 \, GeV$. It was found there that the meson masses go to zero
at $T_c$ regardless of this mass.

A simple hydrogenic model derived from Bethe-Salpeter
equation~\cite{pilkuhn} is invoked in \cite{PLB} to show why the
meson masses are zero at $T_c$ irrespective of $m_q$. The
resulting equation is essentially the Klein-Gordon one since the
spin-dependent interactions can be neglected because of the large
inertial parameter in the denominator of the magnetic moment as
discussed below. It may very well be that this parameter goes to
$\infty$ as $T$ comes down to $T_c$ from above as a result of
confinement, making our approximation and the Weinberg mended
symmetry~\cite{weinberg} exact. In this model the heavy quark
Coulomb potential is started from zero at $r=0$ and increases to
$2 m_q$ (with string breaking ignored) at large distances. The
chirally restored mesons would have energy $\sim m_q$ in this
approximation (half of the absolute value of the $-2 m_q$
potential energy). Introduction of the velocity-velocity
interaction brings the meson masses to zero.

The chirally restored mesons have been found in quenched LGS at $T
\approx 1.4 T_c$ and $1.9 T_c$ \cite{AHN,PP}. In the heavy quark
approximation Park {\it et al.} \cite{PLB} deduced that the
binding energy was $0.15 \, GeV$ and the thermal mass $m_q \sim
1.2 \, GeV$ at $1.4 \, T_c$. Nonlinearities were not displayed;
all 32 modes were degenerate. On the other hand, Brown {\it at
el.} \cite{BLR} found in the light quark formalism that just above
$T_c$ there were large nonlinearities. The width for $\rho \to \pi
\pi$ was estimated as $380 \, MeV$. (Above $T_c$ this is the only
interspecies transition, the $\pi$ and $\sigma$ and the $\rho$ and
$a_1$ being equivalent). Possibly the nonlinearities produced
enough noise in the lattice calculations to prevent them from
being extended below $1.4 T_c$. (Difficulties did appear
\cite{H}). In any case the large $\Gamma (\rho \to \pi \pi)$
signals strong interspecies interactions.

\sect{Rescaling at $T_c$.} Just above $T_c$, the color singlet
(Coulomb) potential predominates completely over the colored
interactions. Thus only the colorless states matter. This can be
understood roughly in the following way.

As long as hadrons with masses are present, chiral symmetry is
broken and scales are set by the chiral symmetry breaking one
$4\pi f_{\pi} \approx 1\, GeV$, where $f_{\pi} \approx 90 \, MeV$
is the pion decay constant. However, once chiral symmetry is
restored at $T_c$, the scale is determined by the $\pi$, $\sigma$,
$\rho$, $a_1$ meson (scalar) masses which are zero at $T_c$.
Suppose we describe this effect -- which is clearly absent in
perturbation theory and hence highly non-perturbative -- in terms
of an $effective$ Coulomb interaction with an ``$effective$
charge" that simulates movement towards the infrared, so that the
``Coulomb interaction," evaluated in lattice gauge calculations of
the Polyakov loop \cite{b1}, becomes large, $\alpha_s \sim 2$. The
effective coupling constant, with the 4/3 value of the Casimir
operator and the factor 2 for velocity dependence included, comes
to $\alpha_s \to \sim 16/3$ as $T\to T_c$. Since $\alpha_s = g^2/4
\pi$ this means that $g \to \sim 8$, and $g > 1$ being the strong
coupling. This strong coupling is manifested in the interactions
for $T \geq T_c$.

At first sight it may seem strange that the masses of the bound
states which result from lattice calculations employing quite
large bare (or current) quark masses $\bar{m}$ all drop so sharply
toward zero. That explicit chiral symmetry breaking plays an
insignificant role can be understood by realizing that in the
presence of the explicit breaking, the total mass is
\begin{equation}
M = \sqrt{m^{2}_{th} + \bar{m}^{2}} \approx m_{th} + { \bar{m}^{2}
\over 2 m_{th}} \label{Mass}
\end{equation}
with $m_{th} \sim 1 \, GeV$, so that $M \sim m_{th}$ unless
extremely high $\bar{m}$'s are used in the LGS. Thus, the lattice
calculations of the system entropy at $T_c$ depends little on the
explicit chiral symmetry breaking.

\sect{As the soft glue melts the interactions go to zero.} Let us
now go up to $T_c$ from below in the hadron sector. Since the
hadronic interactions are given by the exchange of soft gluons, as
these ``melt" and disappear, the interactions go to zero. We have
not yet rigorously established but we believe that the melting of
the soft glue corresponds to the gauge coupling $g_V\to 0$, i.e.,
``the vector manifestation (VM)" in HLS theory~\cite{HY}. In fact
both the vector coupling constant $g_V$ and the mass $m_{\rho}$ go
to zero at a fixed point at $T_c - \epsilon$. This gives rise to
``hadronic freedom." This implies that as the fireball expands and
$T$ drops below $T_c$ in heavy-ion processes, interactions between
the hadrons cease, building up again as $T$ drops to zero.

In going up to $T_c$ from below, the dynamically generated masses
of the mesons, such as the $\rho$, go to zero leaving only the
bare mass $\bar{m} \sim 5 \, MeV$. Since the width of the $\rho$
goes as \cite{BLR} $ {\Gamma^{*}_{\rho} \over \Gamma_{\rho}} \sim
\left( {m^{*}_{\rho} \over m_{\rho}} \right)^{5}$ replacing
$m^{*}_{\rho}$ by $5 \, MeV$ one sees that the effects from the
explicit chiral symmetry breaking are negligible. Thus, in
conjunction with eq.~(\ref{Mass}), we see that the explicit chiral
symmetry breaking changes the description of the chiral
restoration transition only negligibly from that in the chiral
limit; {\it i.e.} the effects from crossover, rather than second
order phase transition would be difficult to see. We might have
expected this from the fact that the behavior of the glue (see
Fig.2 of \cite{PLB}), which has no quarks in it, is a good guide
to the chiral restoration transition.

We should note that whereas the VM with movement toward
$g^{*}_{V}=0$ at the fixed point brings the vector mesons to zero
mass just below $T_c$ and the symmetry determining the zero mass
of $\pi$ and $\sigma$ just above $T_c$ is chiral symmetry (which
protects the pion from mass), the apparent $SU(4)$ symmetry which
seems to set $m^{*}_{\rho}$ and $m^{*}_{a_{1}}$ equal to $m_{\pi}$
just above $T_c$ is not exact. The magnetic moments of quarks and
antiquarks above $T_c$ are given by eq.~(21) of Brown {\it et al.}
\cite{BLRS}
\begin{equation}
\mu_{q, \bar{q}} = \pm {\sqrt{\alpha_s} \over p_0}
\end{equation}
where $p_0 = E-V$, and $-V \sim 2 \, m_{th} \sim 2 \, GeV$. The
large inertial mass in the denominator of the spin dependence
suppresses it greatly, but since the spin dependence is not zero,
the $SU(4)$ symmetry is not exact.

\sect{Concluding Discussion.} Although our principal focus was on
the vicinity of $T_c$, it is interesting to extend the scenario to
higher $T$. The four detector groups conducting research at RHIC
have announced that they have created a new form of matter, ``the
perfect liquid''. The initial temperature at which RHIC matter is
formed is $\sim 2 T_c$. This liquid follows from the scenario of
Shuryak and Zahed \cite{SZ2} which can easily be understood from
the results of \cite{PLB}. Namely, the thermal masses, divided by
$T$, decrease toward the perturbative value
\begin{equation}
{m_{th} \over T} ={ g \over \sqrt{6}} \, ,
\end{equation}
so that the $q \bar{q}$-pairs will become unbound at $T \sim 2
T_c$. As they unbind their scattering amplitudes become very large
in magnitude, going through $\pm \infty$, so that the quark and
antiquark mean free paths are small, giving the low viscosity of
the perfect liquid.

Now there can be myriads of other states, both colored and
colorless, which contribute substantially to the pressure as $T$
increases to $\gsim 1.2 - 1.3 \, T_c$ \cite{SZ2}. Admixture with
these states will broaden the colorless modes we considered above,
spreading the width but leaving the total strength of our
colorless collective modes intact. These can be seen in the heavy
quark calculations of Asakawa {\it et al.} \cite{AHN} as broad
peaks of widths $\sim 3 \, GeV$ at energy $\sim 4 \, GeV$ for $T
\approx 1.9 T_c$.

From the perfect liquid as temperature drops the system goes
into that of colorless bound states as we have outlined. Near
$T_c$, colored objects are too massive to figure in
thermodynamics. Then, as $T$ drops through $T_c$, a complete
``hadronic freedom" sets in as the interactions between mesons
(other than pions) vanish. Later the on-shell hadrons get their
interactions back.

Although there is evidence that the vector-meson mass drops in
density~\cite{SB,rho}, the
consequence of VM on which our principal thesis is based will be
the cleanest near the critical point in that while the mass drops
to zero, the width is to vanish more rapidly. Lattice measurement
should be able to verify or falsify this prediction. Theoretical
issues that need to be studied are the effect of quark masses
ignored in the RGE flow to the VM fixed point and the mechanism as
to how the same massless mesons go across between $T_c\pm
\epsilon$. The latter may be addressed by holographic dual QCD
once one knows how to calculate corrections to the large 't Hooft
limit.

 \vskip -0.1cm
 \sect{Acknowledgments} We thank
%the Brookhaven LGS group
F. Karsch, P. Petreczky, O. Kaczmarek and F. Zantow for the
generous sharing of their calculations. We are grateful to H.-J.
Park and C.-H. Lee who, with G.E.B. carried out the calculations
of the bound state masses in \cite{PLB} using their results. GEB
and BAG were partially supported by the U.S. Department of Energy
under Grant No. DE-FG02-88ER40388 and part of the work of MR was
supported under Brain Pool program of Korea Research Foundation
through KOFST, grant No. 051S-1-9..

\vskip 0.2cm

$\bullet$ {\bf Note added}

\vskip 0.2cm After the paper was accepted for publication, it was
brought to our attention that we had inadvertently left out an
important paper that appeared in the same year as that of
Ref.\cite{YD} by Su Houng Lee~\cite{SHLEE} who suggested the notion
of two glues exploited in our paper and discussed also the
possibility of nonperturbative quark-gluon dynamics being operative
above $T_c$.

%\vskip 0.5cm

\end{document}